\documentstyle[12pt,epsfig]{article}
\newcommand{\beq}{\begin{equation}}
\newcommand{\eeq}{\end{equation}}
\newcommand{\ber}{\begin{eqnarray}}
\newcommand{\eer}{\end{eqnarray}}

\def\fun#1#2{\lower3.6pt\vbox{\baselineskip0pt\lineskip.9pt
  \ialign{$\mathsurround=0pt#1\hfil##\hfil$\crcr#2\crcr\sim\crcr}}}

\title{\huge Leading Logarithms in Field Theory}
\author{ 
{Ugo Aglietti, Guido Corb\`o}
 \\
$~~~$\\
Dipartimento di Fisica, Universit\` a di Roma  \lq La Sapienza\rq  \\ 
INFN, Sezione di Roma I, P.le A. Moro 2, 00185 Roma, Italy 
\\
$~~~$\\
{Luca Trentadue}
\\
$~~~$\\
Dipartimento di Fisica, Universit\`a di Parma \\
INFN Gruppo Collegato di Parma, 43100 Parma, Italy 
\\}

\date{}

\begin{document}

\maketitle

\vskip 2.0cm

\begin{abstract}
\noindent 

We consider the Sudakov form factor in effective theories and we 
show that one can derive correctly the double logarithms of the original, 
high-energy, theory. We show that in effective theories it is possible to 
separate explicitely soft and hard dynamics being these two regimes
related to velocity conserving and to velocity changing operators
respectively.
A new effective theory is sketched which extracts the leading
collinear singularities of the full theory amplitudes. 
Finally, we show how all leading logarithmic effects
in field theory can be obtained by means of simple effective 
theories, where they correspond to a renormalization effect.

\end{abstract}

\newpage

\section{Introduction}
\label{introduction}

The treatment of logarithmic singularities in local 
field theories is well understood. Logarithmic ultraviolet 
divergencies are treated with the renormalization procedure.
In particular regions within the range of the momenta
of the particles, however, additional logarithmic corrections
do appear in the perturbative expansion. These additional infrared 
logarithms come from the region of small-momenta of the amplitudes
and have therefore a dynamical origin. The appearence of these 
logarithms has been analyzed in detail long ago \cite{suda}
in the context of perturbative evaluation of form factor in  
Quantum Electrodynamics. These logarithms are the usually called Sudakov 
double logarithms.  

In this work we propose a new way to look at the Sudakov double logarithms 
in field theory by considering effective theories.
Effective theories are a powerful tool to describe complicated
dynamical processes. They are simpler than the original theories
and give the same predictions in a given energy range.
In general, to reproduce the whole dynamics of the original theory
it is necessary to use a tower of effective theories, covering
interval by interval the entire energy scale.
In this paper we extremize this idea.
We use two effective theories: 
the Heavy Quark Effective Theory $(HQET)$ \cite{geo}
and the Large Energy Effective Theory $(LEET)$ \cite{dgr}.

The paper is organized as follows.
In sec.\ref{sudff} we show that all double logarithms of 
Sudakov form factors can
be extracted by a combined use of the Heavy Quark Effective
Theory and the Large Energy Effective Theory.  
We also note that the Sudakov double logarithms in the effective theory 
do not represent a dynamical effect but are simply a 
{\it renormalization}.
They are associated with the renormalization constants 
of operators of dimension three which change the velocity 
of the effective quark. These operators have the form
\beq\label{dim3}
O(x)~=~h^{\dagger}_{v'}(x)~h_v(x).
\eeq
We consider in sec.\ref{hands} velocity changing
operators of dimension four which, by covariance, have the
form
\beq\label{dim4}
O'(x)~=~h_{v'}^{\dagger}(x)D_{\mu}h_v(x).
\eeq
These operators are related to hard $QCD$ interactions, while
the operators in eq.(\ref{dim3}) are related to hard {\it external}
interactions such as $\gamma$-exchange or weak decays.
The effective theory provides a natural and explicit
separation of soft and hard dynamics, associated 
respectively with velocity conserving
and velocity changing interactions.

In sec.\ref{eft} we describe a new effective theory 
which allows the extraction of leading collinear logarithms,
which we call Collinear Effective Field Theory.
It is a variant of the Large Energy Effective Theory, in which
spin fluctuations are taken into account.

In sec.\ref{summa} we show that all leading logarithmic
effects in field theory can be extracted by means of simple
effective theories where they correspond to a renormalization
effect. 

Sec.\ref{concl} contains the conclusions and an outloook at
future developments. 

In appendix A we evaluate the Altarelli-Parisi kernel $P_{qq}$ in the
Collinear Effective Field Theory.

\newpage

\section{Sudakov Form Factors}
\label{sudff}

In this section we discuss the Sudakov form
factors in the effective theories and their relation with those ones
in the full theory. We consider the $(HQET)$ \cite{geo}
and the $(LEET)$ \cite{dgr} effective theory.
The propagator of the $HQET$, which we may call 
{\it massive eikonal}, is given by
\beq\label{two}
iS_v(k)~=~\frac{1+\hat{v}}{2}~
\frac{i}{v\cdot k+i\epsilon},~~~~~v^2~=~1,
\eeq
while the propagator of the $LEET$, which may be called 
{\it massless eikonal,} is given by
\beq\label{one}
iS_n(k)~=~\frac{\hat{n}}{2}~
\frac{i}{n\cdot k+i\epsilon},~~~~~n^2~=~0
\eeq
(see sec.\ref{eft} for a derivation). The interaction vertices are
derived noting that
\ber
-igt_a~\frac{1+\hat{v}}{2}~\gamma_{\mu}~\frac{1+\hat{v}}{2}
&=&-igt_av_{\mu}~\frac{1+\hat{v}}{2},
\\
-igt_a\frac{\hat{n}}{2}~\gamma_{\mu}~\frac{\hat{n}}{2}
&=&-igt_an_{\mu}~\frac{\hat{n}}{2}.
\eer
It is therefore possible, as far as $QCD$ interactions are
concerned, to omit the spin structure of the effective
propagators,
\ber
iS_v(k)&\rightarrow&\frac{i}{v\cdot k+i\epsilon},
\\
iS_n(k)&\rightarrow&\frac{i}{n\cdot k+i\epsilon},
\eer
and use simplified vertices of the form
\ber
V_v&=&-igt_a v_{\mu},
\\
V_n&=&-igt_a n_{\mu}.
\eer
In general, the massive propagator  extracts the infrared
singularity of the amplitudes, while
the massless propagator extracts the leading infrared times
collinear singularity. The difference originates from the fact that
in the massless
case the `scale' $v^2$ is missing: we send $v^2\rightarrow 0$,
so the $\log v^2$, vanishing in the massive theory, 
becomes singular in the massless one.

There are three possible effective theory form factors, for:
\begin{enumerate}
\item[$(a)$]
light to light transition $(ll)$, $n^2=0,~n'^2=0$;
\item[$(b)$]
heavy to heavy transition $(hh)$, $v^2=1,~v'^2=1$;
\item[$(c)$]
heavy to light transition $(hl)$, $v^2=1,~n^2=0$.
\end{enumerate}
The case of the light to heavy transition is identical to $(c)$.

In general, the effective theory form factors are logarithmically
divergent in the ultraviolet region 
due to the absence of quadratic terms in energy denominators.
This has to be contrasted with the case of the full theory 
form factors, which, for example in the case of the vector or the axial
vector current are ultraviolet finite
because of conservation or partial conservation of the current.
Even in the cases in which the full form factors are
ultraviolet divergent, the $UV$ divergence has a complete
different meaning in the two theories \cite{pzw,aca}, 
as will be clear from the discussion. 
We work throughout the paper at the one-loop order.

Ultraviolet divergences of
effective theory amplitudes are regulated with 
Dimensional Regularization $(DR)$ \cite{dr},
while soft divergences are regulated with off-shell external
states or with a non-vanishing gluon mass $\lambda>0$
\footnote{Technically, it is easier to regulate soft divergences of
massless lines with a virtuality of the external state, while massive
lines are most conveniently regulated with a gluon mass.}.
This implies that the unit of mass $\mu$ has to be considered
as an ultraviolet scale; it would be replaced by the ultraviolet
cut-off $\Lambda_{UV}$ if we were dealing with the bare effective
theory: $\mu\sim\Lambda_{UV}$.

For simplicity's sake, let us consider a vector form factor
in the full theory; we could consider an axial vector current
as well.
In this way we do not need a regulator for the ultraviolet region.
We regulate soft divergences of full theory form factors
with $DR$. $\mu$ therefore has to be considered
as an infrared scale, such as a light parton mass or the
virtuality of external states.
We will see that this complementary use of $DR$ does not cause
any confusion once the physical meaning of the regulators is
understood.
Again for simplicity let us consider the form factors in
the space-like region; in this way we avoid imaginary contributions to form 
factors (the well known $i\pi$ terms), which are irrelevant in this contest. 

\subsection{Massless Case}

Let us start with the simplest case, the on-shell form factor
of a {\it massless} quark \cite{sud,gubpc}:
\ber\label{easy}
&&F(Q^2)~=~
\nonumber\\
&=&\gamma_{\mu}\Bigg[1-2~\frac{C_F\alpha_S}{4\pi}~
\left(\frac{Q^2}{4\pi\mu^2}\right)^{-\epsilon}
\frac{\Gamma(1+\epsilon)\Gamma(1-\epsilon)^2}{\Gamma(1-2\epsilon)}~
\frac{1}{1-2\epsilon}~
\Bigg(\frac{1}{(-\epsilon)^2}+\frac{1}{2}
\frac{1}{-\epsilon}+1\Bigg)~\Bigg]
\nonumber\\
&\simeq&\gamma_{\mu}\Bigg[1-2~\frac{C_F\alpha_S}{4\pi}~
\left(\frac{Q^2}{4\pi\mu^2}\right)^{-\epsilon}
\frac{\Gamma(1+\epsilon)\Gamma(1-\epsilon)^2}{\Gamma(1-2\epsilon)}~
\Bigg(\frac{1}{\epsilon^2}~+\frac{3}{2}~\frac{1}{\epsilon}\Bigg)+\ldots\Bigg]
\nonumber\\
&\rightarrow&\gamma_{\mu}~
\Bigg[1-\frac{C_F\alpha_S}{4\pi}\Bigg(
\log^2\frac{Q^2}{4\pi\mu^2}~-3~\log\frac{Q^2}{4\pi\mu^2}\Bigg)~\Bigg].
\eer
where 
\beq
q~=~p'-p
\eeq 
is the momentum transfer and 
\beq\label{Q2}
Q^2~=~-q^2~=~2p\cdot p'~>~0
\eeq
is the virtual mass squared (the hard scale of the process). 
$\epsilon=2-n/2$, and $n$ is the space-time dimension. 
$C_F=(N^2-1)/(2N)=4/3$ in $QCD$
is the Casimir operator in the fundamental representation of $SU(N)$.
We see the appearance of $1/(-\epsilon)$ poles, of infrared nature.

In the last line of eq.(\ref{easy}) only the terms dependent on the
momentum transfer $Q^2$, which are finite 
for $\epsilon\rightarrow 0$, have been kept.

It is interesting to look also at the expression
for the form factor when
the $\hat{k}$ terms ($k$ is the gluon
momentum) in the numerators of Dirac propagators
have been omitted:
\ber
&&F'(Q^2)~=~
\nonumber\\
&=&\gamma_{\mu}\Bigg[~1-2~\frac{C_F\alpha_S}{4\pi}~
\left(\frac{Q^2}{4\pi\mu^2}\right)^{-\epsilon}
\frac{\Gamma(1+\epsilon)\Gamma(1-\epsilon)^2}{\Gamma(1-2\epsilon)}~
\frac{1}{(-\epsilon)^2}~\Bigg]
\nonumber\\
&\rightarrow&\gamma_{\mu}~
\Bigg[~1-\frac{C_F\alpha_S}{4\pi}~\log^2\frac{Q^2}{4\pi\mu^2}~\Bigg].
\eer

We see that the leading double $1/\epsilon^2$ pole has the
same coefficient in $F(Q^2)$ and $F'(Q^2)$, 
while the simple $1/\epsilon$ pole has a different coefficient
in the two cases: it is $3/2$ in $F$ and $0$ in $F'$.
The blocks of $\Gamma$-functions are identical.
As a consequence, the double collinear logarithm,
$\log^2 Q^2/\mu^2$, is the same in the two cases, while the
single collinear logarithm, $\log Q^2/\mu^2$, has a different
coefficient. This effect occurs also when one considers 
the form factor of a particle with a different spin like the gluon.
With respect to the quark form factor the structure of the leading double 
logarithms remains unchanged provided that $C_F\rightarrow C_A$. On the 
contrary the coefficient of the single logarithms, of collinear origin,
changes reflecting the presence of a particle with a different spin \cite{cet}.
We conclude that the single collinear logarithm is sensitive
to the spin fluctuations: we will see that the same
phenomenon occurs in the $LEET$ and this is the key point
in the construction of an effective theory to extract collinear
logarithms (see sec.\ref{eft}). 

Let us consider now the effective vertex $(a)$,
in which the quarks are taken massless:
\ber\label{five}
F_{ll}&=&
1-2~\frac{C_F\alpha_S}{4\pi}~
\left(\frac{4\pi\mu^2 n\cdot n'}{2~n\cdot p~n'\cdot p'}\right)^{\epsilon}~
\Gamma(1+\epsilon)^2~\Gamma(1-\epsilon)~
\frac{1}{\epsilon^2}
\nonumber\\  
&\rightarrow&1-\frac{C_F\alpha_S}{4\pi}~\Bigg[
\log ^2\left(\frac{2~\mu n\cdot \mu n'}{n\cdot p~n'\cdot p'}\right)
+O\Bigg(\log\frac{2~\mu n\cdot \mu n'}{n\cdot p~n'\cdot p'}\Bigg)\Bigg]
\eer
Infrared and collinear divergences are regulated by taking the 
effective quarks off-shell, $n\cdot p\neq 0,~n'\cdot p'\neq 0$.
There is a double $1/\epsilon^2$ pole, of ultraviolet nature.

Note that the effective diagram does not contain any scale
(unlike the full theory vertex which depends on $Q^2$),
so the theory can develop a single mass ratio and
consequently a single kind of logarithm.

The double logarithm of the full theory, $\log^2 Q^2/\mu^2$,
is reproduced by the effective vertex $(a)$,
\beq
\log^2\frac{Q^2}{\mu^2}~~\iff~~
\log ^2\left(\frac{2~\mu n\cdot \mu n'}{n\cdot p~n'\cdot p'}\right).
\eeq
We interpret the quark momenta as 
\beq
p~=~\mu n,~~p'~=~\mu n',
\eeq 
and we match the infrared regulator $\mu$ of the full theory amplitude
with the off-shellness of the effective theory lines
\beq
n\cdot p,~~n'\cdot p'~~\sim~~\mu
\eeq
To sum up, the massless form factor of the full theory
is correctly reproduced to Double
Logarithmic Approximation $(DLA)$ by the effective vertex $(a)$.

It is interesting to note that the blocks of $\Gamma$-functions entering
in $F$ (or $F'$) and in $F_{ll}$ do coincide up to first order included,
i.e. up to subleading terms:
\beq
\frac{\Gamma(1+\epsilon)\Gamma(1-\epsilon)^2}{\Gamma(1-2\epsilon)}~
\simeq~
\Gamma(1+\epsilon)^2~\Gamma(1-\epsilon)~\simeq~\Gamma(1+\epsilon)~
=~1-\gamma_E\epsilon+O(\epsilon^2)
\eeq
where $\gamma_E\simeq 0.57$ is the Euler constant.

\subsection{Massive Case}
\label{massivo}

Let us consider now the on-shell form factor of a {\it massive} quark
\cite{lan}:
\ber\label{complete}
&&F(Q^2)~=~
\nonumber\\
&=&\gamma_{\mu}~-\frac{C_F\alpha_S}{4\pi}\Bigg[~ 
\gamma_{\mu}\Bigg(
\frac{1+\xi^2}{1-\xi}\Bigg(\log^2\frac{1}{\xi}-\frac{\pi^2}{3}
-4 Li_2(-\xi) + 4\log\frac{1}{\xi}\log(1+\xi)\Bigg)
\nonumber\\
&-&3~\frac{(1+2/3~\xi+\xi^2)}{1-\xi^2}~\log\frac{1}{\xi}
+2~\Bigg(\frac{1+\xi^2}{1-\xi}\log\frac{1}{\xi}
-1\Bigg)\log\frac{m^2}{\lambda^2}\Bigg)
\nonumber\\
&&~~~~~~~~~~~~~~+2~\sigma^{\mu\nu}\frac{q_{\nu}}{m}
\frac{\xi}{1-\xi^2}\log\frac{1}{\xi}~\Bigg]
\eer
where $Li_2(z)$ is the dilogarithm function defined by \cite{grad}
\beq
Li_2(z)~=~-\int_0^z\frac{\log(1-x)}{x}~dx~
=~\sum_{n=1}^{\infty}~\frac{z^n}{n^2},
\eeq
and $\xi$ is defined by the relation
\beq
\frac{(1-\xi)^2}{\xi}~=~\frac{Q^2}{m^2}.
\eeq
We have regulated (for convenience) the infrared divergence 
with a gluon mass $\lambda>0$.
Radiative
corrections generate structures not present in the tree level
form factor (the $\sigma_{\mu\nu}$-term), unlike to what 
happens in the massless case. 

The asymptotic behaviour of the form factor for
\beq
Q^2~\rightarrow~\infty,
\eeq
since
\beq
\frac{1}{\xi}~\sim~\frac{Q^2}{m^2},
\eeq
is obtained taking the limit
\beq
\xi~\rightarrow~0^+.
\eeq
We select terms according to the degree of singularity in this limit.
The leading terms are the double logarithms, $\log^2 1/\xi$, 
the sub-leading terms are the single logarithms,  $\log
1/\xi$, the further sub-leading terms are constants. 
Powers of $\xi\sim m^2/Q^2$ in this framework are 
exponentially small contributions
(the so-called power-suppressed corrections), because
\beq
\xi~=~e^{-\log 1/\xi},~~and~~
\log\frac{1}{\xi}~\rightarrow~\infty.
\eeq
We have:
\ber\label{pff}
F(Q^2)~=~\gamma_{\mu}\Bigg[1-\frac{C_F\alpha_s}{4\pi}
\Bigg(&+&\log^2\frac{Q^2}{m^2}~
+~2~\log\frac{Q^2}{m^2}~\log\frac{m^2}{\lambda^2}+
\\
&-&3~\log\frac{Q^2}{m^2}~-~2~\log\frac{m^2}{\lambda^2}~-\frac{\pi^2}{3}~
+{\rm (pow.~corr.)}\Bigg)~\Bigg]
\nonumber
\eer
where, for completeness, also the subleading logs and the finite term
have been written.
The magnetic form factor is suppressed by a power of $\xi$
and therefore can be neglected to logarithmic accuracy.

The two double logarithms in eq.(\ref{pff}) can be extracted by means of
two effective theories.
The `semihard' double logarithm, $\log^2 Q^2/m^2$
(it is related to loop momenta $k$ in the range $m^2<k^2<Q^2$),
is reproduced by the effective vertex $(a)$ in
which the quarks are taken massless, 
\beq
\log^2\frac{Q^2}{m^2}~~\iff~~
\log ^2\left(\frac{2~\mu n\cdot \mu n'}{n\cdot p~n'\cdot p'}\right).
\eeq
We interpret the quark momenta as for the massless case before,
\beq
p~=~\mu n,~~~~~p'~=~\mu n',
\eeq 
and the quark mass $m$ as the infrared cutoff in the effective theory
$(a)$
\beq
n\cdot p,~~n'\cdot p'~~\sim~~m.
\eeq

The `semisoft' double logarithm, $\log Q^2/m^2\log m^2/\lambda^2$,
is reproduced by the vertex $(b)$, 
in which the quark is treated as an infinite mass particle
\cite{rus,geo2}:
\ber\label{htoh}
F_{hh}&=&1-\frac{C_F\alpha_S}{4\pi}~
2\left(\frac{4\pi\mu^2}{\lambda^2}\right)^{\epsilon}~\Gamma(1+\epsilon)~
\frac{1}{\epsilon}~\Big(v\cdot v'~r(v\cdot v')-1\Big)~ 
\nonumber\\
&\rightarrow&1-\frac{C_F\alpha_S}{4\pi}~2\Big(v\cdot v'~r(v\cdot v')-1\Big)~ 
\log\frac{\mu^2}{\lambda^2}
\nonumber\\
&\simeq&1-\frac{C_F\alpha_S}{4\pi}\Bigg[
2~\log(2v\cdot v')~\log\frac{\mu^2}{\lambda^2}~
-2\log\frac{\mu^2}{\lambda^2}~+\ldots\Bigg]
\eer
where $r(x)=1/\sqrt{x^2-1}~\log(x+\sqrt{x^2-1})$ is the usually
called cusp anomalous dimension \cite{rus}
or velocity dependent anomalous dimension \cite{geo2}.
We have regulated the infrared divergence with a gluon mass $\lambda$,
as in the full amplitude. 
There is a {\it simple} $1/\epsilon$ pole, of ultraviolet nature,
unlike the case $(a)$, because collinear singularities are absent.
We have the correspondence:
\beq
\log\frac{Q^2}{m^2}~\log\frac{m^2}{\lambda^2}~~\iff~~
\log(2v\cdot v')~\log\frac{\mu^2}{\lambda^2}
\eeq
According to eq.(\ref{Q2}) we have indeed:
\beq
\frac{Q^2}{m^2} ~\sim~ 2 v\cdot~v'
\eeq
We identify the renormalization scale $\mu$ of the effective vertex
(or, equivalently, the ultraviolet cutoff $\Lambda_{UV}$ of the
bare effective theory) with the mass in the original vertex:
\beq
\mu^2~\sim~m^2.
\eeq

Therefore we see that the whole form factor can be reproduced
in $DLA$ by means of two effective theories, each one 
describing a different dynamical range.
It is to remark that the quark mass $m$ acts as an 
{\it infrared} cutoff for the {\it massless} effective vertex $(a)$, 
\beq
k^2~<~m^2~~~~(HQET), 
\eeq
while it acts as an {\it ultraviolet} cutoff for the 
{\it massive} effective vertex $(b)$,
\beq
k^2~>~m^2~~~~(LEET).
\eeq
We see a realization of the idea of the tower of effective
theories, each one describing a different energy range,
discussed in the introduction: the $HQET$ describes dynamics
below the quark mass, while the $LEET$ describes dynamics above
the quark mass.

Let us comment about subleading one-loop effects.
The single infrared logarithm,
$\log\mu^2/\lambda^2$, is also reproduced by the effective   
vertex $(b)$ \cite{rus}:

\beq
\log\frac{m^2}{\lambda^2}~~\iff~~\log\frac{\mu^2}{\lambda^2}.
\eeq

The extraction of the other one-loop subleading effect,
the single collinear logarithm, $\log Q^2/m^2$,
cannot be performed with the $LEET$ or the $HQET$, i.e. with the
propagators in eq.(\ref{two}) or (\ref{one}). 
A new, more complicated, effective theory is required (see
sec.\ref{eft}).
We end up this section by observing that the double and the 
single collinear logarithms are the same in the massless and massive full
theory form factors,
\beq
\log^2\frac{Q^2}{m^2}-3\log\frac{Q^2}{m^2}~~
\iff~~
\log^2\frac{Q^2}{4\pi\mu^2}-3\log\frac{Q^2}{4\pi\mu^2},
\eeq
provided one makes the identification $4\pi\mu^2= m^2$
\footnote{In the massive case there are power corrections of the form
$(m^2/Q^2)^n$, which are absent in the massless case.}.

\subsection{Massive to Massless Case}

Let us consider now the form factor for the transition of a quark with
mass $m$ and momentum $p$ with $p^2=m^2$ into a massless quark with
momentum $p'$ with $p'^2=0$ \cite{akh}. 
Following ref.\cite{gubpc} we have:
\ber 
&&F(Q^2)~=~
\nonumber\\
&=&\gamma_{\mu} - 
\frac{C_F\alpha_S}{4\pi}\left(\frac{M^2}{4\pi\mu^2}\right)^{-\epsilon}
\Gamma(1+\epsilon)\Bigg[\gamma_{\mu}\Bigg(\frac{1}{1-2\epsilon}
(1-\xi)F(1+\epsilon,1,1-\epsilon,\xi)\frac{1}{(-\epsilon)^2}
\nonumber\\
&&~~~~~~~~~~+\frac{1-2\epsilon}{(1-\epsilon)}
F(\epsilon,1,2-\epsilon,\xi)~\frac{~1}{-\epsilon}~
-\frac{3}{2}~\frac{1-2/3~\epsilon}{1-2\epsilon}~\frac{~1}{-\epsilon}~
\nonumber\\
&-&2\frac{1}{(1-\epsilon)(1-2\epsilon)}~F(1+\epsilon,1,2-\epsilon,\xi)
\Bigg)~
+\frac{p_{\mu}}{m}~
\frac{-1}{1-\epsilon/2}~F(1+\epsilon,1,3-\epsilon,\xi)
\nonumber\\
&&~~~~~~~~~~~~~~+\frac{{p'}_{\mu}}{m}~\Bigg(~
\frac{-1}{(1-\epsilon)(1-\epsilon/2)}~F(1+\epsilon,2,3-\epsilon,\xi)
\nonumber\\
&&~~~~~~~~~~~~~~+\frac{1}{4}~
\frac{1}{(1-\epsilon)(1-2\epsilon)}~F(1+\epsilon,1,2-\epsilon,\xi)~
\Bigg)~\Bigg]
\eer
where 
\beq
\xi~=~\frac{q^2}{m^2}~=~1-\frac{2p\cdot p'}{m^2}
\eeq
and $F(a,b,c,z)=~_2F_1(a,b,c;z)$ 
is the hypergeometric function, defined
by the series \cite{grad}:
\beq ~_2F_1(a,b,c;z)~=~
1+\frac{a~b}{1~c}z+\frac{a(a+1)~b(b+1)}{1~2~c(c+1)}z^2
+\ldots
\eeq 
$DR$ regulates the collinear divergence related to the massless line
and the infrared divergences, i.e. $\mu$
can be thought of as the light quark mass or the gluon mass.
Radiative corrections produce new structures
with respect to the tree level (the $p_{\mu}$ and $p_{\mu}'$ terms), 
as it happens in the massive case.

There are two different dynamical situations, according to the
condition
\ber
(\alpha)~~~~~p\cdot p' &\gg& m^2;
\\
(\beta)~~~~~p\cdot p' &\ll& m^2.
\eer
To understand the physical meaning of these inequalities, 
let us take the massive quark at rest,
$p=(m,\vec{0})$.
Condition ($\alpha$) means $E\gg m$, i.e. a light quark energy $E$
much larger than the heavy quark mass $m$. This means that
a large energy $E-m=E(1-\epsilon),~\epsilon\ll 1$, is `pumped' 
inside the system by the current.
In case $(\alpha)$ we have:
\beq 
F(Q^2)~\simeq~\gamma_{\mu}~
\Bigg[1-\frac{C_F\alpha_S}{4\pi}\Bigg(
\log^2\frac{2 p\cdot p'}{4\pi\mu^2}~
-\frac{1}{2}\log^2\frac{m^2}{4\pi\mu^2}+\ldots\Bigg)~\Bigg]
\eeq
Due to the increased difficulty, let us limit ourselves
to study the dependence of the form factor on the 4-velocities, 
i.e. we drop the logarithms squared of the mass ratio 
$m/\mu$ \cite{pzw} (these, together with subleading two-loop
effects, will be treated in a future work \cite{fut}):
\beq
F(Q^2)~=~\gamma_{\mu}
\Bigg[1-\frac{C_F\alpha_S}{4\pi}
\log^2 2v\cdot n~+(mass~logs)~\Bigg]
\eeq
Since the momentum transfer $Q^2=2p\cdot p'-m^2$ is much larger
then the heavy quark mass squared $m^2$,
\beq
Q^2~\simeq~ 2 p\cdot p'~\gg~m^2,
\eeq
we can consider both quarks massless. The double logarithm is
reproduced by the effective vertex $(a)$ in eq.(\ref{five}),
\beq
\log^2 2v\cdot n~~~\iff~~~\log^2 2n\cdot n',
\eeq
with the (natural) identification
\beq
v\cdot n~\sim~n\cdot n'.
\eeq
Condition $(\beta)$ means on the contrary $E\ll m$, i.e. most of the
rest energy of the heavy quark $m-E=m(1-\epsilon),~\epsilon\ll 1$, is
absorbed by the external current. We have in this case:
\ber
F(Q^2)&\simeq&\gamma_{\mu}~
\Bigg[1-\frac{C_F\alpha_s}{4\pi}\Bigg(
\log^2\frac{2 p\cdot p'}{4\pi\mu^2}~
+\log^2\frac{2p\cdot p'}{m^2}~
-\frac{1}{2}\log^2\frac{m^2}{4\pi\mu^2}~\Bigg)~\Bigg]
\nonumber\\ \label{diffic}
&=&\gamma_{\mu}
\Bigg[1-\frac{C_F\alpha_S}{4\pi}~
2~\log^2 2v\cdot n~+(mass~logs)~\Bigg]
\eer
Note the additional factor two of case $(\beta)$ with respect to case
$(\alpha)$ in the coefficient of the logarithm squared of the
4-velocities product.

The double logarithm in the last line of eq.(\ref{diffic}) can be
extracted by means of the effective vertex
$(c)$, in which the initial quark is treated as an infinite mass
quark while the final quark is taken as an effective massless quark:
\ber\label{htol}
F_{hl}
&=&1-\frac{C_F\alpha_S}{4\pi}~
\left(\frac{4\pi\mu^2(v\cdot n)^2}{ n\cdot p^{~2} }\right)^{\epsilon}
\Gamma(1+\epsilon)\Gamma(1+2\epsilon)\Gamma(1-2\epsilon)~
\frac{1}{\epsilon^2}
\nonumber\\
&\rightarrow&1-\frac{C_F\alpha_S}{4\pi}~
\frac{1}{2}~\log^2\left(\frac{\mu^2(v\cdot n)^2}{ n\cdot p^{~2} }\right)
\nonumber\\
&=& 1-\frac{C_F\alpha_S}{4\pi}~
2~\log^2 v\cdot n +(mass~logs)
\eer
There is a double, ultraviolet $1/\epsilon^2$ pole coming from the
product of the infrared singularities with the collinear singularity
related to the massless line. 
We have regulated the soft divergences taking only the
massless quark off-shell, $n\cdot p\neq 0$ (this is indeed
sufficient to render the integral finite for $\epsilon< 0$). 
The power of $\mu$ (which is 2)
does not match the number of velocities (which is 4),
so it is impossible to associate a power of
$\mu$ with each 4-velocity, $n$ and $v$, as we did in case $(a)$.

As we said, we considered for illustrative purposes the case of a
vector current, i.e. a conserved current.
In the case of a general form factor, the ultraviolet divergence
also can be extracted with an effective theory \cite{aca}:
we just set to zero all the masses and external momenta in the
full diagrams and we integrate from the $UV$ cutoff 
$\Lambda_{UV}$ down to the largest scale in the diagram
(it can be either a mass or a momentum transfer).

\subsection{Dynamical Logarithms and Renormalization Logarithms}
\label{dlrl}

The effective lagrangian can be written in a covariant form
introducing a field for each velocity as \cite{geo}
\beq\label{leff}
{\cal L}(x)~=~
\int \frac{d^3v}{2v_0}~h^{\dagger}_v(x)~i v\cdot D~h_v(x)~
+~\int \frac{d^3n}{2n_0}~h^{\dagger}_n(x)~i n\cdot D~h_n(x).~~~~~
\eeq
${\cal L}(x)$ contains all the operators diagonal in the velocity
with dimension $d$ less or equal to the
canonical one $n=4$ 
(the field dimensions are $d_h=3/2$ and $d_A=1$ and
mass operators of the form 
$h_v^{\dagger} h_v,~h_n^{\dagger} h_n$ can be set equal to zero
\cite{neu}).

All the operators $O(x)$ with dimension
\beq
d~=~3
\eeq
can be written in a covariant form as
\beq\label{d3}
O(x)~=~\sum_{i,j}^{1,2}
\int \frac{d^3v_i}{2v_i^0}\frac{d^3v_j}{2v_j^0}~ 
c(v_i\cdot v_j)~h^{\dagger}_{v_i}(x) h_{v_j}(x),
\eeq
where we defined for notational simplicity $v_1=v,~v_2=n$.
The effective theory form factors $(a),~(b)$ and $(c)$
correspond formally to the renormalization constants
$c(v_i\cdot v_j)$ of the operators in eq.(\ref{d3}) \cite{rus}.

If we define the renormalization constants as
\ber
c_R(v\cdot v')&=&\frac{c_B(v\cdot v')}{Z_{hh}},
\nonumber\\
c_R(n\cdot n')&=&\frac{c_B(n\cdot n')}{Z_{ll}},
\nonumber\\
c_R(v\cdot n)&=&\frac{c_B(v\cdot n)}{Z_{hl}},
\eer
we have in the $MS$ scheme (according to eqs.(\ref{five}), 
(\ref{htoh}) and (\ref{htol})):
\ber
Z_{hh}&=&1~-~2~\frac{C_F\alpha_S}{4\pi}~\frac{1}{\epsilon}~
\Big(v\cdot v' r(v\cdot v') -1\Big),
\nonumber\\
Z_{ll}&=&1~-~2~\frac{C_F\alpha_S}{4\pi}~\frac{1}{\epsilon^2}+\ldots,
\nonumber\\
Z_{hl}&=&1~-~\frac{C_F\alpha_S}{4\pi}~\frac{1}{\epsilon^2}+\ldots.
\eer
Note that the light-to-light and the heavy-to-light renormalization
constants contain a {\it double} ultraviolet logarithm at the
one-loop level, a novel feature in field theory (in $DR$ we have
that $1/\epsilon_{UV}\rightarrow\log\Lambda^2 _{UV}/\mu^2$, where 
$\mu$ is the renormalization point). 
The coefficient of the double pole in the heavy-to-light vertex
is one-half of that one in the light-to-light vertex because 
the collinear singularity is not coupled to the heavy leg.
  
Therefore we have that the double logarithms of the full theory,
dynamical logarithms, correspond to renormalization logarithms of 
local operators of dimension three in the effective theory.

We argue that subleading terms can be
extracted by means of a combined use of the $HQET$ and a
new effective theory (the collinear effective field theory 
$(CEFT)$) to be discussed in the next section. 
The justification of this conjecture is the following.
As it is well known, subleading effects involve 
both the {\it single infrared} logarithm, $\alpha_S\log m^2/\lambda^2$,
and the {\it single collinear} logarithm $\alpha_S\log Q^2/m^2$
\cite{trentadue}. The former is well reproduced by the $HQET$ 
as we have seen in sec.\ref{massivo} \cite{rus}, while the
latter is well reproduced by the $CEFT$, as we show explicitly
in a particular case in the appendix A.

\section{Hard and Soft Dynamics in the Effective Theory}
\label{hands}

It is natural to consider all the operators of dimension less
or equal to the canonical one $n=4$.
The operators of dimension three have already been considered
in the previous section: they are related to hard external
interactions (such as $\gamma$-exchange or weak decays).
The gauge invariant operators of dimension
\beq
d~=~4,
\eeq
can be written in a covariant form as
\beq\label{d4}
O(x)~=~\sum_{i,j}^{1,2}
\int \frac{d^3v_i}{2v_i^0}\frac{d^3v_j}{2v_j^0}~ 
d_{\mu}(v_i,v_j)~h^{\dagger}_{v_i}(x) D^{\mu}h_{v_j}(x)
\eeq
These operators have a clear physical meaning:
they correspond to `cusps' in space-time associated with
`hard' gluon emission or absorption, i.e. a gluon with momentum 
\beq 
k~\sim~m,~E.
\eeq 
For example, the operator $h_v^{\dagger}~A_{\mu}~h_{v'}$
is associated with the absorption of a gluon with momentum
$k=m(v'-v)$. 

The effective theory provides a natural 
separation of soft and hard dynamics: these two regimes 
are related with the velocity conserving interactions
in eq.(\ref{leff}) and with the velocity changing operators
in eq.(\ref{d4}) respectively. In other words, 
an arbitrary $QCD$ diagram can be decomposed in
effective theory diagrams, in which soft interactions are 
described  by the vertices in eq.(\ref{leff}),
while hard interactions are described by the external 
operators in eq.(\ref{d4}).
This decomposition requires the introduction of a  
scale $\mu$ separating hard momenta from soft momenta,
similar in spirit to the factorization scale of
Wilson's operator product expansion \cite{ope}.
In general a $QCD$ process contains a few hard subprocesses 
while it contains an arbitrary number of soft subprocesses.
To soft processes, factorized with respect to the hard ones,
then the resummation procedure can be also applied in order
to obtain reliable perturbative results \cite{cat}.

This difference is encoded in the effective theory
because the interaction lagrangian, inserted any
number of times, describes soft gluons, while 
external operators, inserted just a few times, describe
hard effects. 

\section{An Effective Theory for Collinear Logarithms}
\label{eft}

We have seen in sec.\ref{sudff} that the Sudakov region 
can be extracted with the $HQET$ and the $LEET$.
As it is well known, $DLA$ Sudakov form factors are related to both
{\it soft} and {\it quasi-collinear} emissions \cite{dks}:
\beq\label{sdr}
\theta~\ll~1,~~~k~\ll~E,
\eeq
where $\theta$ is the gluon emission angle and $E$ and $k$
are the energies of the parent quark and the gluon respectively.

Collinear singularities, related to configurations with
\beq\label{clr}
\theta~\ll~1,~~~~k~\sim~E,
\eeq
can also be extracted 
by means of an effective theory, a new effective theory
which we call Collinear Effective Field Theory 
$(CEFT)$, and whose elements are sketched in the following lines. 
In the propagator of a massless quark,
\beq
iS_F(P)~=~i\frac{\hat{P}}{P^2+i\epsilon},
\eeq
we separate the basic momentum from the fluctuation according to 
\beq
P~=~En+k, 
\eeq
so that
\ber\label{start}
iS_F(En+k)&=&
\Bigg(\frac{ \hat{n} }{2}+\frac{ \hat{k} }{2 E}\Bigg) 
\frac{i}{n\cdot k+k^2/2E+i\epsilon}.
\nonumber\\
&=&\frac{ \hat{n} }{2}~
\frac{i}{n\cdot k+i\epsilon}~+~O\left(\frac{1}{E}\right).
\eer
$n^{\mu}$ is a light-like vector, $n^2=0$,
normalized in such a way that $v\cdot n=1$, where $v$
is a reference time-like vector, $v^2=1,~v_0>0$. 
$E$ is the primordial (large) energy of the quark.
There are two terms of order $1/E$:
\beq
\frac{k^2}{2E}~~~and~~~\frac{\hat{k}}{2E}.
\eeq
They are subleading in the limit $E\rightarrow\infty$
and of the same order in $1/E$ but, as we are going to show, 
they have different dynamical meanings.
The term $k^2/(2E)$, in the denominator, describes both
antiquark excitations and quark transverse momentum dynamics.
We may write 
\beq
k^2~=~(k_0-k_z)(k_0+k_z)-{\vec{k}_T}^2.
\eeq
Without any loss of generality we may take $n=(1;0,0,1)$.
We have, close to the quark mass-shell, 
or equivalently at lowest order in $1/E$, $k_0\sim k_z$. 
If we neglect antiquark creation, a process
very far from the quark mass-shell, we may simplify dynamics
according to
\beq\label{kt}
k^2~\sim~-{\vec{k}_T}^2.
\eeq
Keeping the term $k^2/(2E)$ in the approximation (\ref{kt}),
we account for a transverse motion  (relative to $n$)
of the effective quark.
The transverse motion is important, for example, in bound state
dynamics \cite{aco}. 

The term $k^2$ is small {\it both} in the infrared region and
in the collinear region. In the infrared region all the components
$k_{\mu}$ are small, so the square is small,
while in the collinear region the individual
components $k_{\mu}$ are large but $k_0\sim\mid\vec{k}\mid$,
so that $k^2\sim 0$. Therefore we can neglect the term $k^2/(2E)$
completely for the extraction of the infrared as well as the collinear
region. This is a huge semplification of the dynamics:
the propagator does not involve a sum over all the possible
trajectories of the particle (the Feynman's sum over hystories),
but only the classical path. Just for this reason,
for example, the propagator of the $LEET$
can be computed in closed form in an arbitrary
gauge field:
\beq
iS_n(x\mid 0)~=~\frac{ \hat{n} }{2}~\int d\tau~\delta^{(4)}(x-n\tau)~ 
P~e^{ i\int_0^x A_{\mu} dx^{\mu} }.
\eeq
It is remarkable that both the leading infrared singularites
and the leading collinear singularities are related to such a simple
semiclassical dynamics.
 
Let us now turn to the correction $\hat{k}/(2E)$ related to spin
fluctuations of the quark. This term can be neglected in the 
infrared region in which all the components $k_{\mu}$ are small,
but cannot be neglected in the collinear region, in which the
components are large and are projected over an arbitrary direction
$q_{\mu}$.

In the case of a massive quark the decomposition of the momentum $P$
is done according to the formula
\beq
P~=~mv+k,
\eeq
so we end up with a propagator of the form
\ber
iS_F(mv+k)&=&
\Bigg(\frac{1+\hat{v}}{2}+\frac{ \hat{k} }{2m}\Bigg) 
\frac{i}{v\cdot k+k^2/2m+i\epsilon}.
\nonumber\\
&=&\frac{ 1+\hat{v} }{2}~
\frac{i}{v\cdot k+i\epsilon}~+~O\Bigg(\frac{1}{m}\Bigg).
\eer
In this case the collinear singularity is absent, so this
propagator can extract only infrared logarithms.
It is therefore consistent to neglect both kinds of $1/m$ corrections
as far as logarithmic effects are concerned.
To summarize, an effective propagator for the collinear region
is derived neglecting {\it only} the term $k^2/(2E)$ in the 
denominator of eq.(\ref{start}):
\beq\label{ffn} 
i\tilde{S}_F(k;E)~=~\Bigg(\frac{ \hat{n} }{2} 
+\frac{\hat{k}}{2E}\Bigg)~
\frac{i}{n\cdot k+i\epsilon}.
\eeq
Unlike the $LEET$ case (cf. eq.(\ref{one})), we {\it keep} 
the fluctuating momentum $k$ in the numerator in order to 
account for collinear but not soft configurations.
The propagator (\ref{ffn})
does still depend on the hard scale $E$, contrary to what
happens to the $LEET$ propagator (\ref{one}).
If the quark emits a longitudinal momentum fraction
$1-x$, such that $k\sim -(1-x)En$,
the propagator after the emission looks like
\beq
i\tilde{S}_F(k;E)~
\rightarrow~x~\frac{ \hat{n} }{2}~\frac{i}{n\cdot k+i\epsilon},
\eeq
and takes correctly into account the longitudinal momentum loss
(the Sudakov region (\ref{sdr}) corresponds to the
limit $x\rightarrow 1$). 
In the appendix we describe a sample computation with the $CEFT$,
the evaluation of the non-singlet Altarelli-Parisi kernel.

\section{Leading Logarithms in Quantum Field Theory}  
\label{summa}

Let us end up discussing what seems to us a general 
property of quantum field theories.
We observe that all leading logarithmic effects in field theory
can be extracted by means of a simple effective theory, where they 
correspond to a renormalization effect (i.e.
they are not dynamical, they renormalize a 
parameter of the effective theory).
To justify this statement, we just need to enumerate 
the known sources of logarithmic effects in field theory.
\begin{itemize}
\item Sudakov region (see eq.(\ref{sdr})). 
It has been discussed
at lenght in the main body of the paper and
can be extracted by means of the $HQET$ and the $LEET$;
\item Collinear region (see eq.(\ref{clr})). 
The discussion has been mostly qualitative; 
it requires the introduction of a new
effective theory, the $(CEFT)$, with the propagator given in
eq.(\ref{ffn}); 
\item Infrared region, $k\ll E$.
Infrared singularities of form factors
have been extracted and factorized 
with massive eikonal lines, i.e. with the $HQET$, by Korchemsky and 
Radyushkin in ref.\cite{rus}, so we
refer to these papers for the proof and for a full
discussion; 
\item  Hard region, $k^2\sim \Lambda^2$, where $\Lambda$
is the ultraviolet cutoff. 
According to Wilson's Renormalization Group Transformation \cite{wil}, 
ultraviolet logarithmic divergences 
induce a renormalization
of parameters of the (leading) effective hamiltonian.
\end{itemize}

\section{Conclusions}
\label{concl}

The main conclusion of our work is that leading logarithmic
effects in field theory do have a very simple dynamical origin.
They do not involve the full complexity of quantum field theory
and can be extracted by means of effective theories.
The latter represent very simple kinematical configurations:
particles which move with constant velocity along segments
of a given broken line.
We stressed the difference between the massive eikonal propagator
and the massless eikonal propagator, and that 
the replacement of a Dirac propagator with one of these propagators
depends on the energy scale under consideration.
We have here discussed the various approximations involved
in the construction of an effective theory.
The quadratic term $k^2$ in energy denominators can be neglected
as far as the extraction of the soft region
(i.e. infrared e/o collinear) is concerned.
Neglecting $k^2$ implies a dramatic simplification of the dynamics:
quantum fluctuations are suppressed and particles move along 
classical trajectories only.
We have seen that this term is responsible both for pair creation
and for transverse momentum dynamics.
It cannot be neglected completely, but it can be simplified
($k^2\rightarrow -\vec{k}_T^2$) in the construction
of an effective theory describing bound state dynamics
\cite{aco}, i.e. effects which are somehow complementary to the
logarithmic ones. 
For the collinear region the spin fluctuation ($\hat{k}$) in
numerators of Dirac propagators has to be kept, while
it can be neglected for the extraction of the infrared region.

We believe that the main conclusions of our analysis remain
true when higher order corrections are taken into account.

$~~$

\centerline{\bf Acknowledgments}

$~~$

We wish to thank G.~Martinelli and V. Lubicz for discussions.

$~~$

\newpage

\appendix
\section{Derivation of the Non-Singlet Splitting Function}
\label{apar}

As a sample computation with the $CEFT$ we present 
in this appendix the
derivation of the non-singlet splitting function $P_{qq}$
entering the Altarelli-Parisi equation.

We consider the collision of a massless quark $q$ with momentum
$p_{\mu}$ with $p^2=0$, with a space-like photon 
with momentum $q_{\mu}$, $q^2<0$:
\beq
q~+~\gamma^*~\rightarrow~q~+~X,
\eeq
where $X=0$ or $g$ in the elastic channel or in the inelastic channel 
of order $\alpha_S$ respectively.
Given $p$ and $q$ it is possible to construct two light-like vectors:
$p$ itself and $\eta=xp+q$, where $x$ is the Bjorken variable,
$x=-q^2/s$, and $s=2~p\cdot q$.
The tree $QCD$ diagram (the leading contribution to the
elastic process), gives a rate:
\beq
L~=~2\pi~\delta((p+q)^2)~
\frac{1}{2}~Tr\;{\cal O}~=~\frac{\pi}{s}~\frac{\delta(1-x)}{x}~Tr\;{\cal O}
\eeq
where we defined
\beq
Tr\;{\cal O}~=~Tr\Big[x\hat{p}~\gamma_{\rho}~(x\hat{p}+q)~\gamma_{\sigma}\Big].
\eeq
The elastic process is used in the effective theory to define 
the vertex, i.e. the cusp angle from the kinematics.
We replace the electromagnetic vertex of the full theory
$\overline{q}(x)\gamma_{\mu}q(x)$, with an effective vertex 
of the form $\overline{h}_{n'}(x)\gamma_{\mu}h_n(x)$, 
where $h_n$ and $h_{n'}$ are fields of the $CEFT$. 
As the only possible choice, we take:
\beq 
En~=~p,~~E'n'~=~\eta.
\eeq
Let us consider now single gluon emission to order $\alpha_S$.
We take the gluon propagator in planar gauge with the gauge
vector along $\eta$, so that the gluon polarization sum is
\beq
S_{\mu\nu}(k,\eta)~=~-g_{\mu\nu}
+\frac{ \eta_{\mu}k_{\nu}+\eta_{\nu}k_{\mu} }{ \eta\cdot k }.
\eeq
With this gauge choice, collinear singularities are decoupled
from the final quark leg, and the only relevant diagram is the
(one-rung, cut) ladder diagram \cite{dks,ua}.
The ladder diagram is given in the full theory by
\beq
F~=~N~\int d^4 k~\frac
{\delta((p-k)^2)~\delta((k+q)^2)}{(k^2)^2}~X_{\rho\sigma},
\eeq
and in the effective theory by
\beq
E~=~N~\int d^4 k~\frac
{\delta((p-k)^2)~\delta(2\eta\cdot(k-xp))}{(2~p\cdot(k-p))^2}~
X_{\rho\sigma},
\eeq
where $N=\alpha_S C_F/(2\pi)$ is a normalization factor and
$X_{\rho\sigma}$ is defined as
\beq
X_{\rho\sigma}~=~Tr\Big[ \hat{p}~\gamma_{\nu}~\hat{k}~\gamma_{\sigma}~
(\hat{k}+\hat{q})~\gamma_{\rho}~\hat{k}~\gamma_{\mu}\Big]~
S^{\mu\nu}(k,\eta).
\eeq
It encodes the effects of the spin structure of the gluon and 
of the quark for the process. 
$X_{\rho\sigma}$ is the same in full $QCD$ and in the
$CEFT$ because the latter does not involve any approximation
for the quark spin structure, but simply a different writing
of the helicity sum (see eq.(\ref{ffn})).

The $\delta$-function of the emitted gluon is clearly the same in the
two theories (we did nothing to the gluon field), 
while the $\delta$-function of the final quark 
and the propagator of the virtual quark are different.
In this particular kinematical configuration however, since 
the gluon is on the mass shell, $(p-k)^2=0$,
the quark propagators in the two theories actually coincide:
$k^2= (p+(k-p))^2 = 2 p\cdot(k-p)$.

We use the Sudakov decomposition of the loop momentum $k$,
\beq
k~=~\alpha~\eta~+~\beta~p~+~k_T,
\eeq
where $k_T$ is the transverse momentum, orthogonal both to
$p$ and $\eta$: $k_T\cdot p=k_T\cdot\eta=0$.
For notational simplicity, let us define $k_T^2=\vec{k}_T^2$.
The loop measure is 
\beq
d^4k~=~\frac{s}{2}~d\alpha~d\beta~d^2k_T.
\eeq
We use the notation of ref.\cite{dks} and we refer to this book
for further details and for a physical discussion of $DIS$
in full $QCD$.
In terms of the Sudakov variables the full diagram reads
\beq
F~=~N\frac{s}{2}\int d\alpha~d\beta~d^2k_T~
\frac
{\delta(\alpha(1-\beta)s+k_T^2)~\delta(s(1+\alpha)(\beta-x)-k_T^2)}
{(k_T^2/(1-\beta))^2},
\eeq
while the effective diagram reads
\beq E~=~N
\frac{s}{2}\int d\alpha~d\beta~d^2k_T~
\frac
{\delta(\alpha(1-\beta)s+k_T^2)~\delta(s(\beta-x))}
{(k_T^2/(1-\beta))^2}.
\eeq
The only difference is the argument of the $\delta$-function
of the final quark:
\ber
QCD:&&s~(1+\alpha)~(\beta-x)-k_T^2,
\nonumber\\
CEFT:&&s~(\beta-x).
\eer
It is well known that the collinear singularity (the collinear log)
is related to the quasi-real gluon emission,
so that
\beq
\beta~\sim~x~\sim~1,~~\alpha~\ll~1,~~\frac{k_T^2}{s}~\ll~1.
\eeq
In this region the two integrands look the same.
More explicitly, if we perform the integrations over $\alpha$ and
$\beta$, we are left with identical expressions as far as the collinear
singularity is concerned:
\beq
F~=~E~=~
\frac{N}{2s}\int \frac{d^2 k_T}{k_T^2}~(1-x)~\frac{X_{\rho\sigma}}{k_T^2}
+(terms~with~no~coll.~log.).
\eeq 
It is to note that in full $QCD$ one makes a set of approximations
to extract the logarithmic structure. In the $CEFT$ these approximations
are `automatic' in the sense that they are alreadt built-in in the
lagrangian.
The evaluation of $X_{\rho\sigma}$ gives
\beq\label{trace1}
X_{\rho\sigma}~=~2~k_T^2~\frac{1+x^2}{x~(1-x)^2}~Tr\;{\cal O}~+~O(k_T^4)
\eeq
Summing the tree diagram with the effective diagram
of order $\alpha_S$, we recover the famous one-loop expression 
for the parton density:
\beq
L~+~E~=~\Bigg[\delta(1-x)~+~\frac{C_F\alpha_S }{2\pi}~
P_{qq}~\log\frac{Q^2}{\mu^2}\Bigg]~\sigma_0
\eeq
where $\sigma_0$ is the hard cross section in the lowest order,
$\sigma_0=\pi/(x~s)~Tr\;{\cal O}$ and 
\beq\label{vero}
P_{qq}~=~\frac{1+x^2}{1-x}.
\eeq

It is interesting to compute $X_{\rho\sigma}$ also in the $LEET$.
We have in this case:
\ber
X_{\rho\sigma}^{(LEET)}&=&Tr\Big[ \hat{p}~\gamma_{\nu}~\hat{p}~
\gamma_{\sigma}~\hat{\eta}~\gamma_{\rho}~\hat{p}~\gamma_{\mu}\Big]~
S^{\mu\nu}(k,\eta)
\nonumber\\
&=&2~k_T^2~\frac{2}{x(1-x)^2}~Tr\;{\cal O}.
\eer
The latter expression coincides 
with $X_{\rho\sigma}$ in eq.(\ref{trace1})
in the limit $x\rightarrow 1$, i.e. in the Sudakov region.
The $LEET$ kernel is
\beq\label{kerleet}
P_{qq}^{LEET}~=~\frac{2}{1-x},
\eeq
so that the difference with the `right' $CEFT$ (or $QCD$) kernel is
\beq
P_{qq}^{(CEFT)}-P_{qq}^{(LEET)}~=~
\frac{1+x^2}{1-x}-\frac{2}{1-x}~=~-(1+x)
\eeq
The integral of the difference over $x$ gives:
\beq
\int_0^1 dx~(-)(1+x)~=~-\frac{3}{2},
\eeq
i.e. an infrared finite term.
This coefficient accounts for the single collinear logarithm,
$\log Q^2/m^2$.
We also observe that if we take into account the longitudinal
momentum loss in the trace of the $LEET$, i.e. if we make the
replacement $\hat{p}\rightarrow x\hat{p}$ in the numerators of quark
propagators, the trace changes into:
\ber
X_{\rho\sigma}^{(LEET)}&\rightarrow&Tr\Big[ \hat{p}~\gamma_{\nu}~x\hat{p}~
\gamma_{\sigma}~\hat{\eta}~\gamma_{\rho}~x\hat{p}~\gamma_{\mu}\Big]~
S^{\mu\nu}(k,\eta)
\nonumber\\
&=&2~k_T^2~\frac{2x^2}{x(1-x)^2}~Tr\;{\cal O}.
\eer
In the latter case the kernel is
\beq\label{eurist}
\tilde{P}_{qq}~=~\frac{2~x^2}{1-x},
\eeq
i.e. it comes out {\it smaller} then the `true' one (\ref{vero}).
The correct kernel is the average of (\ref{kerleet}) 
and (\ref{eurist}).
This may be interpreted by saying that spin fluctuations are
so important in collinear dynamics that their effect 
cannot be represented with
any kind of effective constant helicity sum, such as $\hat{p}$ or 
$x\hat{p}$.
In this particular case, the `right kernel' (\ref{vero}) is the average
of the $LEET$ kernel (\ref{kerleet}) with the effective kernel
(\ref{eurist}) in which strict longitudinal momentum loss, 
$\hat{p}\rightarrow x\hat{p}$ has been assumed.

To summarize, we have shown that the $CEFT$ reproduces in a very simple
way the kernel $P_{qq}$ of full $QCD$.

\vfill

\end{document}